\documentstyle[twoside,fleqn,espcrc2]{article}
\newcounter{num}

\title{Weak radiative decays of hyperons and of charm and beauty baryons
\thanks{Research supported in part by a grant from the Ministry of
Science and the Arts and by the Fund for Promotion of Research at the Technion.}
}
\author{Paul Singer\\
Department of Physics, Technion-Israel Institute of Technology, 32000
Haifa, Israel}
\begin{document}

\begin{abstract}
A review is presented of the weak radiative decays of baryons.  It
includes an analysis of the possible contributions of electromagnetic
penguins
to these decays, a survey of the difficulties still encountered in the
sector of hyperon decays and a short account on some new developments
on this topic.  The theoretical treatments on charm and beauty baryon
decays are summarized, with a good outlook for their detection.
\end{abstract}

\maketitle
\section{INTRODUCTION}

Although hyperon decays have been under\linebreak  scrutiny for some three decad
the subject still carries the burden of a major puzzle and of discrepancies
between existing data and a variety of theoretical models [1,2].  At the
other end of quarks spectrum, there are no data yet on weak radiative decays
of heavy baryons; however, estimates [3-6] for some of these modes allows
us to anticipate optimistically their future detection.

In the $(s,d,u)$ sector, the interesting weak radiative processes are
two-body decays.
These decays proceed with branching ratios of the order of
$(1-3)\times10^{-3}$, like
$\Sigma^+\rightarrow p\gamma$,
$\Lambda \rightarrow n\gamma,\Xi^0\rightarrow \Sigma^0\gamma$,
or of the order $10^{-4}$, like $\Xi^-\rightarrow\Sigma^-\gamma$
and the expected $\Omega^-\rightarrow\Xi^-\gamma$[7].  The three-body
decays $\Lambda\rightarrow p\pi^-\gamma$,
$\Sigma^{+,-}\rightarrow n\pi^{+,-}
\gamma$ proceed as expected for inner bremsstrahling processes with
branching ratios close to
$10^{-3}$ and are not of our concern here.  On the other hand, the two-body
exclusive heavy baryon weak radiative processes like
$\Lambda_b\rightarrow \Lambda^0\gamma$,
$\Xi^-_b\rightarrow \Xi^-\gamma$ are not necessarily dominating the
radiative channel and as we shall see one expects these modes to be
substantially
smaller than the inclusive ones, e.g. $BR[\Lambda_b\rightarrow X(s)\gamma]
>> BR[\Lambda_b\rightarrow \Lambda^0\gamma]$.  Nevertheless, the study of the
exclusive channels could provide important physical insights.

These weak radiative processes result from an interplay of electroweak and
gluonic interactions.  Presently, their theoretical treatment requires the
inclusion of separate short-distance (SD) and long distance (LD)
contributions
[8,9,10,5].    The
estimate of the relative size of the two types of processes is an issue
to be determined for every specific process.  If, for instance, one is
confident that in a certain process the long-distance emission is a
rather small perturbation, like in
$B\rightarrow X(x)\gamma$, $B\rightarrow K^*\gamma$,
such processes may be assigned the strategic role of testing
the Standard Model
[11,12] as well as the testing of theories beyond it [13].

The next section surveys the possible role of the SD single-quark transition
$Q\rightarrow q\gamma$ in the weak radiative decays of strange,
charm and beauty baryons.

\section{ELECTROWEAK~PENGUINS~IN BARYON RADIATIVE WEAK DECAYS}

At the quark level there are three types of processes which contribute
to the weak radiative decays of baryons, classified [14,15] as single-,
two-, and three-quark transitions.  The two-quark transition
corresponds to $W-$ exchange, with the photon radiated by the participating
quarks, and it is essentially a long-distance process.
The three-quark transition,
where the quark not participating in $W$-exchange radiates a photon,
is strongly suppressed [15].  The single-quark transition involves a SD
contribution due to the electromagnetic (EM) penguin diagrams [10,11,12]
as well as possible LD contributions [16,17].

Before turning to the role of the EM penguins in the weak radiative baryon
decays, one should mention the powerful analysis of Gilman and Wise (GW)[14].
In their paper, GW checked the hypothesis that {\em all} weak radiative
hyperon decays in the 56-multiplet of SU(6) are driven by the single-quark
transition $s\rightarrow d\gamma$.   They
determined the strength  from the $\Sigma^+\rightarrow p\gamma$ decay
and proceeded to calculate from this the expected branching ratios for
$\Lambda \rightarrow n\gamma$,
$\Xi^0\rightarrow\Sigma^0\gamma$,
$\Xi^0\rightarrow \Lambda\gamma$,
$\Xi^-\rightarrow \Sigma^-\gamma$,
$\Omega^-\rightarrow\Xi^-\gamma$ and
$\Omega^-\rightarrow\Xi^{-*}\gamma$.
Their predictions exceed the experimental rates by one or two orders of
magnitude for the various decays.  Thus, the hypothesis that
{\em all these decays proceed via the single-quark transition is untenable}.
However, it must be stressed that the analysis of GW does not preclude
substantial contributions from $s\rightarrow d\gamma$, whether SD
[18,19] or LD [17], in only some of the hyperon radiative decays.

In the standard electroweak model, the flavour-changing $Qq\gamma$
vertex with the
$Q,q$ quarks on the mass-shell has the form
\begin{eqnarray}
\Gamma_\mu &=&
\frac{e}{4\pi^2}\frac{G_F}{\sqrt{2}}(q)
\sum_\lambda V^*_{\lambda Q}V_{\lambda q}
[F_{1,\lambda}(k^2)(k_{\mu} k\hspace{-0.20cm}/ \nonumber \\
&-& k^2\gamma_\mu)
\frac{1-\gamma_5}{2} \nonumber\\
&+& F_{2,\lambda}(k^2)i\sigma_{\mu\nu}k^\nu(m_Q\frac{1+\gamma_5}{2}\nonumber\\
&+& m_q\frac{1-\gamma_5}{2})](Q).    
\end{eqnarray}
$F_1(q^2)$ and $F_2(q^2)$ are the charge radius and magnetic form
factors respectively,
calculated [20] in electroweak theory in terms of masses of quarks and $W$;
$V_{ab}$ are Cabibbo-Kabayashi-Maskawa\linebreak (CKM) matrices.  For $(sd\gamma)$
and $(bs\gamma)$ one has $\lambda=u,c,t$ and for $(cu\gamma)$ the
contribution is from $\lambda = d,s,b$.

The $F_1$ term does not contribute to decays with real photons.  It is,
however, relevant in decays involving leptons like $B\rightarrow X(s)
\ell^+\ell^-$ [21],
$\Sigma^+\rightarrow p\ell^+\ell^-$ [22],
$\Omega^-\rightarrow \Xi^-\ell^+\ell^-$[19].  In this paper we restrict
our discussion to decays with real photons, to which only $F_2$ contributes.

The quantity of physical interest is the $Qq\gamma$ vertex with QCD
corrections.
The effective hamiltonian has the form
\begin{equation}
H_{\rm eff} = - \frac{G_F}{\sqrt{2}}\lambda \sum  C_i(\mu) O_i(\mu)
\end{equation}
where $\lambda$ represents symbolically products of CKM matrices,
 $O_i(\mu)$ is a complete set of dimension-six operators and $C_i(\mu)$
are Wilson coefficients.  Explicit expressions for the strange, charm and
beauty sectors are given in Refs.~[23], [24] and [12] respectively.
$O_{1,2}$ are current-current operators, $O_3-O_6$ are strong penguin
operators and $O_7,O_8$ are magnetic operators, of EM and gluonic
type respectively.  In particular the EM penguin operator required by
Eq.~(1) has the form
\begin{eqnarray}
&&O_7= \frac{e}{8\pi^2} (\bar{q})_\alpha \sigma^{\mu\nu}[m_Q(1+\gamma_5)\nonumber\\
&&~~~+ m_q(1-\gamma_5)](Q)^\alpha F_{\mu\nu} \ .      
\end{eqnarray}

The application of the QCD corrections using the renormalization
group equations
endows (3) with a coefficient $C^{\rm eff}_7$, which is a
linear combination of
$C_i(\mu)$ and has been calculated for all three sectors, at least to
leading order.  We are thus in  a position to determine quantitatively
the contribution of the EM penguin to the baryonic radiative weak decays.

In the strangeness sector, the replacement by the QCD-corrections of a
quadratic GIM cancellation by logarithmic dependence, increases
$F_2$ by about three orders of magnitude [2,10].  The value of $C_7^{\rm eff}
(sd\gamma)$ has been reevaluated recently with better accuracy [17,25].
Using the new value we estimate the SD contribution to the typical pole
decay $\Sigma^+\rightarrow p\gamma$ and to the decays which have been
singled out [18] as potential windows to $s\rightarrow d\gamma$, namely
$\Omega^-\rightarrow\Xi^-\gamma$ and $\Xi^-\rightarrow\Sigma^-\gamma$.
Using wave functions of Ref.~[14] we find
\begin{equation}
\Gamma(\Sigma^+\rightarrow p\gamma)^{\rm SD}_{s\rightarrow d\gamma}
/\Gamma(\Sigma^+\rightarrow p\gamma)_{\rm exp} =
2\times10^{-5}   
\end{equation}
Hence in hyperon radiative decays driven by LD poles the
$s\rightarrow d\gamma$ transition
does not play a noticeable role.  On the other hand, one finds
\begin{eqnarray}
\Gamma(\Omega^-\rightarrow \Xi^-\gamma)^{\rm SD}_{s\rightarrow d\gamma} =
6.4\times10^{-12} {\rm eV}   \ .     
\end{eqnarray}
Using the recently determined [26] upper limit $\Gamma(\Omega^-
\rightarrow\Xi^-\gamma)_{\rm exp} < 3.7 \times 10^{-9}$eV one concludes [17]
that in this decay the amplitude ratio SD/LD is larger than 1/25.
Obviously, this is a remarkable result.

A similar calculation for $\Xi^-\rightarrow\Sigma^-\gamma$ gives
\begin{eqnarray}
\Gamma(\Xi^-\rightarrow \Sigma^-\gamma)^{\rm SD}_{s\rightarrow d\gamma} =
8.3\times10^{-13} {\rm eV} \ ,       
\end{eqnarray}
which indicates a contribution of SD of about 4\% in the amplitude of this
decay.

The transition $c\rightarrow u\gamma$ has been treated in detail, including
QCD corrections, only recently [24].
Contributions from all three quark loops are comparable in size,
like in the strangeness sector.  Likewise, the QCD corrections enhance
also here enormously the transition, leading to a $c\rightarrow u\gamma$
width which is increased by five orders of magnitude.  However, even with
increased strength the $c\rightarrow u\gamma$ EM penguin is too small to
play a role in weak hadronic radiative decays.

The $b\rightarrow s\gamma$ transition has been treated in great theoretical
detail [12,27].  In this case, the contribution of the $t$-quark loop is
strongly dominant so that other contributions are usually omitted.  The
recent measurements by CLEO of $B\rightarrow K^*\gamma$[28] and $B\rightarrow X(
\gamma$ [29] confirm the original expectations [11] that these modes are
dominated by the EM penguin transition $b\rightarrow s\gamma$.
We expect therefore
$b\rightarrow s\gamma$ to play a central role also in beauty baryon decays [5].

Hence, the role of the SD $Q\rightarrow q\gamma$ transition in the
baryonic weak radiative decays is of different nature
in each of the
three sectors: it is totally negligible in the charm sector, it dominates
the appropriate decays in the beauty sector, and
plays a modest role
in some of the hyperon decays like $\Omega^-\rightarrow \Xi^-\gamma$
and $\Xi^-\rightarrow\Sigma^-\gamma$.\\

\section{THE HYPERON SECTOR}

The amplitude for the transition $B(p) \rightarrow B^\prime(p^\prime)\linebreak
 + \gamma(k)$
is
\begin{eqnarray}
&&M(B\rightarrow B^\prime\gamma) = i e G_F \bar{u}(p^\prime)\sigma_{\mu\nu}
(A+ \nonumber \\
&&~~~~ B\gamma_5)\epsilon^\mu k^\nu u (p)     
\end{eqnarray}
where $A(B)$ are the parity-conserving (-violating) amplitudes.
The angular distribution of the decay is characterized by
an asymmetry parameter
$\alpha_h$, given by
\begin{eqnarray}
\alpha_h = 2Re(A^*B)/(|A|^2+|B|^2)       \ .     
\end{eqnarray}
Table 1 summarizes the experimental situation, based on Ref. [7] except
for the entry on $\Omega^-\rightarrow\Xi^-\gamma$ which is based on a new
experiment [26].  The recent analysis on the $\Sigma^+\rightarrow p\gamma$
width based on 31900 events [30], not included in Table 1, gives
$BR(\Sigma^+ \rightarrow p\gamma)=(1.20\pm 0.06 \pm 0.05)\times 10^{-3}$.

A puzzling feature is the large negative asymmetry detected in
$\Sigma^+\rightarrow p\gamma$.  According to Hara's theorem [31], in the
limit of SU(3)-flavour symmetry the PV-amplitudes in $\Sigma^+\rightarrow
p\gamma$ and $\Xi^-\rightarrow\Sigma^-\gamma$ should vanish, causing a
vanishing asymmetry.  Many articles have been devoted to this question as
exemplified by Ref.~[32].  It has also been argued [33] that in a quark
description the Hara theorem does not hold and the problem could lie
in the ``translation'' of the quark basis to the hadronic world.  So
far, there is no convincing explanation for this large SU(3)-breaking.

A large number of models have been construc- ted to treat the processes
of Table 1, most of them attempting a ``unified'' picture for the radiative
hyperon decays.  Among these models, there are pole
models [34], quark models [15], skirmion models [35], Vector Meson Dominance
models [36] and chiral models [37].  In many of these attempts, one
accomplishes firstly a fit to the well measured $\Sigma^+\rightarrow p\gamma$
mode, and predictions are made for other decays, though Refs.~[35],
[37] do
not follow this pattern.  Unfortunately, none of the existing models can
reproduce simultaneously all the features in Table 1.  In fact, comparing
various models (see, e.g. Table 7.1 of Ref.~[1] and Table II of Ref.~[2])
one finds strong disagreements for the yet unmeasured quantities.  In the
following, we restrict ourselves to an analysis of the better understood
physical features in these decays.
\begin{table*}
Table 1\\
\underline{The experimental status of the hyperon radiative decays}

\begin{tabular*}{\textwidth}{@{}l@{\extracolsep{\fill}}ccc}\hline
Decay & Branching Ratio($10^{-3}$) & Asymmetry Parameter\\ \hline
$\Sigma^+\rightarrow p\gamma$ & $1.25\pm0.07$ & $-0.76 \pm 0.08$ \\
$\Lambda\rightarrow n\gamma$ & $1.75\pm0.15$ &                \\
$\Xi^0\rightarrow \Sigma^0\gamma$ & $3.5~\pm0.4$ &$~0.20 \pm 0.32$ \\
$\Xi^0\rightarrow \Lambda^0\gamma$ & $1.06\pm0.16$ & $~0.4~ \pm 0.4$ \\
$\Xi^-\rightarrow \Sigma^-\gamma$ & $0.127\pm0.023$ &\\
$\Omega^-\rightarrow \Xi^-\gamma$ & $< 0.46$ \\
$\Omega^-\rightarrow \Xi^{-*}\gamma$\\ \hline
\end{tabular*}
\end{table*}
The analysis of Section 2 has shown that the contribution of SD emission
is negligible in the four decays proceeding at the $10^{-3}$ level,
namely $\Sigma^+\rightarrow p\gamma$, $\Lambda\rightarrow n\gamma$, $\Xi^0
\rightarrow \Sigma^0(\Lambda^0)\gamma$.  It also can account for only a
fraction of the decays proceeding at the $10^{-4}$ level or lower,
$\Xi^-\rightarrow\Sigma^-\gamma$, $\Omega^-\rightarrow \Xi^-\gamma$,
$\Omega^-\rightarrow\Xi^{*-}\gamma$, as already established for
$\Xi^-\rightarrow \Sigma^-\gamma$[9].  Thus, in all hyperon radiative
decays the LD emission plays the predominant role.  A further dynamical distinct
arises from the valence quark structure of the hyperons and the explicit
form of $H^{\Delta S = 1}_{\rm eff}$ (Eq.~[2]).  For the above group of four
decays $H_{\rm eff}$ induces pole diagrams [e.g. $\Sigma\rightarrow (p,N^*)
\rightarrow p\gamma$, etc.], which dominate over multiparticle intermediate
states.  A suitable combination of the $\frac{1}{2}^+$ baryons and
$\frac{1}{2}^-$ resonance poles can lead to large asymmetries.  However,
the poor knowledge of some of the couplings involved leads to a widely
divergent spectrum of predictions.

The second group of three decays involves particles $\Omega^-(sss)$,
$\Xi^-(ssd)$, $\Sigma^-(sdd)$ with no $u$-valent quark, i.e. there are no
$W$-exchange diagrams to generate poles.  These decays will then proceed
via two-hadron intermediate states.  Gluonic penguins may also contribute;
explicit calculations [38] indicate that such penguin contributions are
considerably suppressed.

Thus, from a dynamical point of view, there are two distinct groups:
the ``pole decays'' $(\Sigma^+\rightarrow p\gamma$, $\Lambda \rightarrow
n\gamma$, $\Xi^0\rightarrow \Sigma^0\gamma$, $\Xi^0\rightarrow\Lambda^0
\gamma$) and the ``non-pole decays'' ($\Xi^-\rightarrow\Sigma^-\gamma$,
$\Omega^-\rightarrow\Xi^-\gamma$, $\Omega^-\rightarrow\Xi^{*-}\gamma$),
which are driven by different mechanisms.  As an example of a ``non-pole''
calculation we mention $\Xi^-\rightarrow\Sigma^-\gamma$ [8],
where the main LD contribution
is due to the $(\Lambda\pi^-$) intermediate state.  The imaginary
part of Eq.~(7) is then
\begin{eqnarray}
Im M(\Xi^-\rightarrow \Sigma^-\gamma) = \frac{1}{2}
\int \frac{d^4k}{(2\pi)^2}
\delta(k^2-m_\pi^2)&\hspace{-0.35cm}\delta[(p- \nonumber \\
k)^2-M^2_\Lambda]M(\Xi^-\rightarrow \Lambda
\pi^-)\cdot T(\pi^-\Lambda\rightarrow\gamma\Sigma^-)&\nonumber\hfill(9)
\end{eqnarray}
giving [9] $Im A^{\rm LD} = 0.94$MeV, $Im B^{\rm LD} = -8.3$ MeV.  For
the real part, dominated by an infrared log divergence in the chiral limit,
one finds $ReA^{\rm LD}\linebreak = 0$, $ReB^{LD} = -6.9$MeV.  Including uncerta
one obtains [9]
\setcounter{equation}{9}
\def\theequation{\arabic{equation}}
\begin{eqnarray}
&&\frac{\Gamma(\Xi^-\rightarrow\Sigma^-\gamma)}{\Gamma(\Xi^-\rightarrow {\rm all})}
= (1.8 \pm 0.4)\times10^{-4}; \nonumber \\
&& \alpha_h(\Xi^-\rightarrow\Sigma^-\gamma)=
-0.13\pm0.07 \ .                                         
\end{eqnarray}
The value for the width agrees well with experiment; the measurement of the
asymmetry is required to confirm the physical picture.

\section{A VECTOR MESON DOMINANCE APPROACH FOR LONG DISTANCE TRANSITIONS
$Q\rightarrow q\gamma$}

A new approach to the calculation of the LD contributions to the radiative
decays $b\rightarrow s(d)\gamma$ has been suggested recently [16] and was
applied to $s\rightarrow d\gamma$ and
hyperon radiative decays in Ref. [17].  The basic idea is to
calculate the LD emission via the $t$-channel, assuming the vector meson
dominance (VMD) of the hadronic electromagnetic current [39].  A hybrid
approach is employed in converting from the nonleptonic hamiltonian
expressed in terms of quark operators to the process $Q\rightarrow qV
\rightarrow q \gamma$.  It should be mentioned that an older ``$s$-channel''
attempt to calculate the LD contribution to $s\rightarrow d\gamma$ [40]
uses a problematic mixture of particles and quarks on equal footing in
intermediate loops.

Let us present the new approach [16,17] by considering the relevant
$O_1,O_2$ operators in the $\Delta S=1$ sector of Eq.~(2)
\begin{eqnarray}
&&H^{\Delta S=1}_{\rm eff} = \frac{G_F}{\sqrt{2}}
\sum_{\eta=u,c,t} V_{\eta s}V^*_{\eta d}(C_{1,\eta} O_{1,\eta}+ \nonumber \\
&&~~~C_{2,\eta}
O_{2,\eta})+H.C.                                    
\end{eqnarray}
\setcounter{num}{12}
\setcounter{equation}{0}
\def\theequation{\arabic{num}\alph{equation}}
\begin{eqnarray}
O_{1,\eta} = \bar{d}\gamma_\mu(1-\gamma_5)\eta_\beta\bar{\eta}_\beta
\gamma^\mu(1-\gamma_5)s               \ ,                          
\end{eqnarray}
\begin{eqnarray}
O_{2,\eta} = \bar{d}\gamma_\mu(1-\gamma_5)\eta\bar{\eta}\gamma^\mu
(1-\gamma_5)s  \ .                                       
\end{eqnarray}
Using factorization, one obtains the amplitude for $Q\rightarrow q V$
proportional to $a_2g_V$, where $\langle V(k)|\bar{\eta}\gamma_\mu
\eta|\linebreak 0\rangle = i g_V(k^2)\epsilon_\mu^+(k)$ and $a_2 = c_1 +
\frac{C_2}{N}$, $N$ being the number of colors.  For the hyperon
decays, the $\eta=t$ contribution is negligible and using
$V_{cs}V^*_{cd} \simeq - V_{us}V_{ud}^*$ and the Gordon decomposition to
extract the transverse part, one has
\setcounter{equation}{12}
\def\theequation{\arabic{equation}}
\begin{eqnarray}
&& A^{s\rightarrow d\gamma}_{\rm LD} = - \frac{eG_F}{\sqrt{2}} V_{cs}V^*_{cd}
a_2(\mu^2)
\left[\frac{2}{3} \sum_i\frac{g^2_{\psi_i}(0)}{m_{\psi_i}^2} -\right. \nonumber
\\
&&~~~\left. \frac{1}{2}
\frac{g^2_{\rho}(0)}{m_{\rho}^2}-\frac{1}{6}
\frac{g^2_{\omega}(0)}{m_{\omega}^2}\right] \ . \nonumber \\
&&\frac{1}{M^2_s-M_d^2}\bar{d}\sigma^{\mu\nu}
(M_sR - M_dL)sF_{\mu\nu}            \ .                
\end{eqnarray}
A phenomenological value $a_2(\mu^2)\geq0.5$ is assumed [17]. $M_s, \ M_d$
are constituent quark masses, $R,L$ projection operators and the
summation covers the six narrow $1^-\psi$ states.

The $\Omega^-\rightarrow\Xi^-\gamma$ decay is calculated [17] from (13) using
the formalism of Ref.~[14].
Using the experimental bound [26] of
$\Gamma_{\rm exp}(\Omega^-\rightarrow\Xi^-\gamma)<3.7\times10^{-9}$eV one
obtains  the relation
\begin{eqnarray}
|C_{\rm VMD}|&=&\left|\frac{2}{3}\Sigma_i
\sum_i\frac{g^2_{\psi_i}(0)}{m_{\psi_i}^2} - \frac{1}{2}
\frac{g^2_{\rho}(0)}{m_{\rho}^2}- \right. \nonumber \\
&&\left. \frac{1}{6}
\frac{g^2_{\omega}(0)}{m_{\omega}^2}\right|
< 0.01 {\rm GeV}^2      \ .                           
\end{eqnarray}
\mbox{The relation (14) represents} a remarkable cancellation at the 30\%
level.  It also determines $\sum_ig^2_{\psi_i}(0)/m_{\psi_i}^2 =
0.045\pm0.016$GeV$^2$, implying a strong $k^2$ dependence in the
$\psi_i-\gamma$ couplings which reduces their value by a factor of
$\simeq 6$ from $k^2=m^2_{\psi_i}$ to $k^2=0$.  This conclusion agrees
well with independent determinations of $g_{\psi_i}(0)$ from
photoproduction and decays [16].

We expect $|C_{\rm VMD}|$ to be quite close to the upper limit value (15),
which in turn implies that $BR(\Omega^-\rightarrow\Xi^-\gamma)$ should
be close to the experimental upper limit of Table 1.
The two-body intermediate
states contribute [8] to the BR of this decay only $0.8\times 10^{-5}$.
The application of this approach to $\Xi^-\rightarrow\Sigma^-\gamma$
gives for the LD contribution to the rate
from $s\rightarrow dV$ an upper limit
of 80\%.  For a pole decay like $\Sigma^+\rightarrow p\gamma$ the same
contribution is less than 1\%.  These values confirm the
consistency of the dynamical picture
discussed in this section.

\section{CHARM BARYON DECAYS}

Charm baryons containing one $c$ quark are usually classified according
to the SU(3) representation of the two light quarks, which can form a
symmetric sextet (with spin 1) or an antisymmetric antitriplet (with
spin 0).  The spin $\frac{1}{2}$ antitriplet is composed of $\bar{B}^3_c
(\Lambda_c^+, \ \Xi_c^+, \ \Xi_c^0$).  The sextet baryons have
spin $\frac{1}{2}$ $(B^6_c)$ or spin  $\frac{3}{2}$ $(B_c^{6*})$.
The particles forming it are
$(\Sigma^{++}_c, \ \Sigma^+_c, \ \Sigma^0_c, \ \Xi^{,+}_c$,\linebreak $\Xi^{,0}_?
\Omega^0_c$).
The $\bar{B}^3_c$ particles and $\Omega^0_c$ decay weakly, while the
rest of sextet particles decay strongly
$(\Sigma_c^{++,+,0}\rightarrow\Lambda_c^+\pi^{+,0,-}$) or electromagnetically\linebreak
$(\Sigma^+_c\rightarrow\Lambda_c^+\gamma$, $\Xi_c^{,+,0}\rightarrow\Xi^{+,0}\gamma$.
In the following, we shall consider only two-body weak radiative decays of
charm baryons.

The SD contribution from $c\rightarrow u\gamma$ to the radiative decays
was shown
to be negligible [24], hence the main mechanism for the decays is
$W$-exchange.
Since the radiative decays are ``cleaner'' than other weak multiparticle
decay channels
of $B_c$ to strongly interacting particles, one may hope that their
estimate will be quite reliable.  We start our considerations by
firstly classifying these decays according to their CKM strength:\\
{\em CKM allowed decays $(\Delta C=\Delta S = -1)$}: $\Lambda^+_c
\rightarrow\Sigma^+\gamma; \ \Xi^0_c\rightarrow\Xi^0\gamma$.\\
{\em CKM forbidden decays $(\Delta C=-1; \ \Delta S = 0)$:} $\Lambda^+_c
\rightarrow p\gamma; \ \Xi_c^+\rightarrow\Sigma^+\gamma;
\Xi^0_c\rightarrow \Lambda(\Sigma^0)\gamma;
\ \Xi^0_c
\rightarrow\Xi^0\gamma$.\\
{\em CKM doubly-forbidden decays $(\Delta C = -\Delta S = -1)$:}
$\Xi^+_c\rightarrow p\gamma; \ \Xi^-_c\rightarrow n\gamma; \ \Omega^0_c
\rightarrow\Lambda^0(\Sigma^0)\gamma$.\\
The photon energy in these decays is considerably larger than in the
hyperon decays, ranging between 833 MeV in $\Lambda_c\rightarrow\Sigma^+\gamma$
to 1124 MeV in $\Omega_c^0\rightarrow\Lambda\gamma$.

Kamal has pioneered [3] this field by calculating $\Lambda^+_c\rightarrow
\Sigma^+\gamma$ from two-quark $W$-exchange bremsstrahlung transitions
of type $c+d \rightarrow s+u+\gamma$.  Summing all relevant diagrams
 one obtains an
effective Hamiltonian which is used to calculate the amplitudes
$A,B$ of Eq.~(7).
Using harmonic oscillator wave functions for the baryons involved, a branching
ratio of nearly $10^{-4}$ is obtained.  Uppal and Verma [4] have improved
the relativistic corrections of this calculation and have also introduced strong
flavour dependence in the harmonic oscillator wave functions.  The results
of their two models, together with an updated value of Ref.~[3] and results
from a heavy-quark effective theory calculation [5] with $c$ and $s$
quarks as heavy are presented in Table 2 for the CKM allowed decays.
\begin{table*}
Table 2.\\
\underline{Theoretical Estimates for Charm Baryon Decays}
\begin{tabular*}{\textwidth}{@{}l|@{\extracolsep{\fill}}c|c|c|c|c|c|c}\hline
 & \multicolumn{4}{c|}{Branching Ratio ($10^{-4}$)} &
\multicolumn{3}{c}{Asymmetry}\\ \cline{2-8}
Decay Mode & Ref.~[3] & Ref.~[4] & Ref.~[4] & Ref.~[5]
& Ref.~[4] & Ref.~[4] & Ref.~[5] \\
 & & I & II & & I & II & \\ \hline
$\Lambda_c\rightarrow\Sigma^+\gamma$ &0.67 & 0.45 & 2.9 & 0.49
& -0.013 & 0.02 & -0.86\\
$\Xi_c^0\rightarrow\Sigma^0\gamma$ & & 0.19 & 1.3& 0.31 & -0.042 & -0.01~ &
-0.86 \\ \hline
\end{tabular*}
\end{table*}

Branching ratios  for the CKM-forbidden decays
$\Lambda^+_c\rightarrow p\gamma$,
$\Xi^+_c\rightarrow \Sigma^+\gamma$, $\Xi^0_c\rightarrow \Lambda\gamma$,
$\Omega^0_c\rightarrow\Xi^0\gamma$ were also estimated in Ref.~[4] and
found to be
generally of the order of $10^{-5}$.

Finally, we comment on the weak radiative decays of heavy baryons with
several $c$
quarks.  Among these, of particular interest is
$\Xi_{cc}^+\rightarrow\Xi^+_c\gamma$
which is CKM allowed and expected with a $10^{-4}$ branching ratio.
There are
also a couple decays which cannot proceed via $W$-exchange.  These are
$\Xi^{++}_{cc}\rightarrow\Sigma^{++}_c\gamma$ and
$\Omega^{++}_{ccc}\rightarrow
\Xi^{++}_{cc}\gamma$, which could be driven by the
$c\rightarrow u\gamma$ transition.
Since the SD contribution is very small, these decays would
constitute a direct window
to the LD $c\rightarrow u\gamma$ process, or possibly to effects beyond the
standard model.

\section{BEAUTY BARYON DECAYS}

As it was explained in Section 2, the SD contribution plays a prominent role
in the $b$-sector.  Therefore, we shall classify the beauty baryon two-body
weak radiative decays as follows: (A) SD decays driven by the EM penguin $b
\rightarrow s\gamma$, which includes $\Lambda^0_b\rightarrow\Lambda^0\gamma$;
$\Lambda_b^0\rightarrow \Sigma^0\gamma$; $\Xi^0_b\rightarrow\Xi^0\gamma$;
$\Xi^-_b \rightarrow\Xi^-\gamma$; $\Omega^-_b\rightarrow\Omega^-\gamma$.
(B) LD decays which are  described on the quark level by two-quark
$W$-exchange transitions
accompanied by photon radiation.  To this group belong
$\Lambda_b^0\rightarrow \Sigma_c^0\gamma$;
 $\Xi^0_b\rightarrow\Xi_c^0\gamma$;
$\Xi^0_b\rightarrow\Xi^{,0}_c\gamma$.
The decays in both groups are CKM doubly-forbidden, the matrix element being
proportional to $V_{tb} V_{ts}^*\sim\lambda^2$ for group (A) and to
$V_{ud}V_{bc}^*\sim \lambda^2$ for group (B).  The photon energies are in
the several GeV range, e.g. $E_\gamma=2.71$GeV for $\Lambda^0_b\rightarrow
\Lambda^0\gamma$.

Theoretical calculations for these decays were performed only recently [5,6].
For group (A) the transition amplitude for
$B_i\rightarrow B_f\gamma$ is given
by the short-distance QCD-corrected $O_7$ operator
\begin{eqnarray}
M(B_i\rightarrow B_f\gamma) = \frac{iG_F}{\sqrt{2}} \frac{e}{4\pi^2}
C^{\rm eff}_7 V_{tb}V_{tb}^*\epsilon^\mu k^\nu&\nonumber \\
\langle\bar{B}_f|\bar{s}\sigma_{\mu\nu}
[m_b(1+\gamma_5)+m_s(1-\gamma_5)]b|B_i\rangle&  
\end{eqnarray}
where $C_f^{\rm eff}=0.31$ [24,27].  The LD contribution to the $b\rightarrow
s\gamma$ transition is estimated to be at the level of a few percent only
[16,17], which allows us to neglect it.  The authors of Ref.~[5] use two
methods to treat the $\Lambda_b\rightarrow\Lambda\gamma$ decay, - the heavy
quark symmetry scheme with both $b$ and $s$ treated as heavy, and the MIT bag mo
In the first method, they obtain for the A,B amplitudes of Eq.~(7)
\begin{eqnarray}
&& A,B = \frac{C^{\rm eff}_7}{4\sqrt{2}\pi^2} V_{tb}V_{ts}^*
\left(1\pm \frac{m_s}{m_b}\right. -
\nonumber \\
&&~~~ \left. \frac{\bar{\Lambda}h}{2m_s}\right)\xi(v\cdot v^\prime)
\end{eqnarray}
where $\xi(v\cdot v^\prime)$ is the Isgur-Wise function and $h$ is a
function of $v\cdot v^\prime$.  Allowing for reasonable variation of the
various parameters involved, Cheng et al. [5] conclude that
\begin{eqnarray}
BR(\Lambda^0_b\rightarrow\Lambda^0\gamma) = (0.5-1.5)\times10^{-5}   
\end{eqnarray}
>From their amplitude, one obtains
$\alpha_h(\Lambda^0_b\rightarrow\Lambda^0\gamma)\linebreak =0.9$.

In the heavy $s$ quark limit, $\Lambda^0$ behaves as an antitriplet
heavy baryon while
$\Sigma^0$ as a sextet heavy baryon.  $\Lambda^0_b$ belongs to an antitriplet.
Accordingly, $b\rightarrow s\gamma$ will not induce in the limiting case
$\Lambda^0_b\rightarrow\Sigma^0\gamma$ which
is a sextet-antitriplet transition
and one is led to
\begin{eqnarray}
\Gamma(\Lambda^0_b \rightarrow\Sigma^0\gamma)<<\Gamma(\Lambda^0_b\rightarrow
\Lambda^0\gamma)                        
\end{eqnarray}
The other decays of group (A) are more difficult to treat (several heavy quarks
baryon).  In any case, branching ratios somewhat smaller than in Eq.~(17)
are expected,
also due to wave function overlap suppression especially in
$\Omega^-_b\rightarrow
\Omega^-\gamma$.

For the transitions of group (B) an effective Lagrangian is constructed [5]
from the diagrams of the $W$-exchange bremsstrahlung processes
$b+u\rightarrow c+d+
\gamma$, $b+\bar{d}\rightarrow c+\bar{u}+\gamma$.  Branching ratios smaller by
at least one order of magnitude than in group (A) are obtained [5],
even if maximal
overlap for the static bag wave functions is assumed:
\begin{eqnarray}
&&BR(\Xi^0_b \rightarrow \Xi^{0}_c\gamma) = 6.4\times10^{-8};
\alpha_h = - 0.47\nonumber \\
&&BR(\Xi^0_b \rightarrow \Xi^{,0}_c\gamma) = 5.7\times10^{-7};
\alpha_h = - 0.98\nonumber \\
&&BR(\Lambda^0_b \rightarrow \Sigma^0_c\gamma) = 1.2\times10^{-6};
\alpha_h = - 0.98\nonumber
\end{eqnarray}

The basic decay mechanism $b\rightarrow s\gamma$ actually leads to a
multitude of exclusive states in the radiative $\Lambda_b$ decay, like
$\Lambda_b\rightarrow\Lambda(1405)\gamma$,
$\Lambda(1520)\gamma$,
$\Lambda(n\pi)\gamma$,
$\Lambda\eta\gamma$,
$\Lambda\eta^\prime\gamma$, etc.
Hence it is of interest to estimate the expected
$\Lambda^0_b\rightarrow X(s)\gamma$ branching ratio and the percentage of it
of the lowest exclusive mode,
$\Lambda^0_b\rightarrow \Lambda^0\gamma$.
We use the measured [29]
$B\rightarrow X(s)\gamma$ to calculate
$\Gamma(b\rightarrow s\gamma) = (1\pm 0.35)\times10^{-7}$eV.
Assuming
$\Gamma(\Lambda^0_b\rightarrow X(s)\gamma)/\Gamma(\Lambda_b\rightarrow
{\rm all})\linebreak
\simeq \Gamma(b\rightarrow s\gamma)/\Gamma(\Lambda_b\rightarrow {\rm all})$
and the measured $\Lambda^0_b$ life-time[7] we estimate
\begin{eqnarray}
\frac{\Gamma(\Lambda^0_b\rightarrow X(s)\gamma)}
{\Gamma(\Lambda_b\rightarrow {\rm all})}
=(1.6\pm0.5)\times10^{-4} \ .                   
\end{eqnarray}
Hence, the calculations presented above lead to
\begin{eqnarray}
\frac{\Gamma(\Lambda^0_b\rightarrow \Lambda^0\gamma)}
{\Gamma(\Lambda_b\rightarrow X(s)\gamma)}
\simeq(6.0\pm3.5)\% .                   
\end{eqnarray}
The figure we obtained is not very different from  the mesonic sector,
where one has [28,29]
$\Gamma(B\rightarrow K^*\gamma)/\Gamma(B\rightarrow X(s)\gamma)=0.2\pm0.1$

An analysis [41] of the angular distribution of the photon in
$\Lambda^0_b\rightarrow
X(s)\gamma$ with polarized $\Lambda_b^0$, using the heavy
quark effective scheme, shows
that deviations from free quark decay are generally small and are
significant mostly for photons
emitted in the forward direction with respect to $\Lambda^0_b$ spin.
However, as a consequence of the functional form of the EM penguin the photons
are emitted preferentially backwards.

\section{CONCLUDING REMARKS}

We highlight here several points, some of which are of direct relevance to
forthcoming and contemplated experimental programmes:

\# As a result of the theoretical activity of last few years, a clear picture
emerges on the the importance of short-distance radiation in the weak
radiative
decays of baryons.  Thus, the electromagnetic penguin
$Q-q\gamma$ (with gluonic
corrections) plays a major role in the beauty sector, dominating processes
like $\Lambda_b\rightarrow X(s)\gamma$,
$\Lambda_b\rightarrow\Lambda\gamma$.
The charm penguin $c\rightarrow u\gamma$ is too weak to play any noticeable
role in charm baryon radiative decays, while the strange penguin
$s\rightarrow
d\gamma$ occupies an intermediate position, contributing to a
possibly detectable
extent in a few hyperon decays ($\Omega^-\rightarrow\Xi^-\gamma$,
$\Xi^-\rightarrow\Sigma^-\gamma$, $\Omega^-\rightarrow \Xi^{*-}\gamma$).

\# The measurements of the rate and asymmetry parameter of
$\Omega^-\rightarrow\Xi^-\gamma$ should be given high priority, since
there is good probability that both SD and LD radiation contributes
measurably to it.  This decay could constitute the main desired window to the
EM penguin in the strangeness sector $s\rightarrow d\gamma$, in addition to
providing interesting information on couplings of vector mesons to photons
from the LD contribution.

\# It is difficult at present to favour any of the  competing models describing
pole hyperon decays like $\Sigma^+\rightarrow p\gamma$, $\Xi^0\rightarrow
\Sigma^0\gamma$, etc.  Since the various models diverge mostly in the
prediction of the asymmetry parameter, good measurements of this parameter in
$\Lambda\rightarrow n\gamma$, $\Xi^0\rightarrow\Lambda\gamma$ and $\Xi^0
\rightarrow \Sigma^0\gamma$ should finally alllow one to resolve the
unsettled situation.

\# The measurement of the asymmetry parameter in the decay $\Xi^-\rightarrow
\Sigma^-\gamma$ will distinguish between the dynamical picture [8,9] for
non-pole decays which leads to Eq.~(10), and alternative mechanisms [34-37].

\# Theoretical estimates indicate that charm baryon CKM allowed radiative
decays will occur with a branching ratio of $\sim 10^{-4}$, making the search
for these decays a realistic proposition.  One expects $BR(\Lambda^+_c\rightarrow
\Sigma^+\gamma) = 1^{+1}_{-0.5}\times10^{-4}$, $BR(\Xi^0_c\rightarrow\Xi^0\gamma
=(0.8\pm0.5)\times10^{-4}$.
The CKM-forbidden decays, like $\Lambda^+_c\rightarrow p\gamma$, $\Xi^0_c
\rightarrow\Lambda(\Sigma^0)\gamma$, $\Xi^+_c\rightarrow \Sigma^+\gamma$,
$\Omega^0_c\rightarrow\Xi^0\gamma$ are expected to occur with branching ratios
of $10^{-5}$ or less.

\# Beauty baryons have detectable weak radiative decays induced by short distanc
electromagnetic penguins.  The inclusive decay $\Lambda_b\rightarrow X(s)
\gamma$ is expected to have a branching ratio of ($1.6\pm 0.5) \times
10^{-4}$.  The most frequent exclusive mode is probably
$\Lambda_b\rightarrow\Lambda\gamma$ expected to occur with a branching ratio
of $(1\pm0.5)\times10^{-5}$.
On the other hand, $\Lambda^0_b\rightarrow
\Sigma^0\gamma$ is expected from heavy quarks symmetry considerations to be
be much smaller.  Radiative decays to charm baryons $\Lambda^0_b\rightarrow
\Sigma^0_c\gamma$, $\Xi^0_b\rightarrow\Xi^0_c\gamma$, $\Xi^0_b\rightarrow
\Xi^{,0}_c\gamma$ are expected in the $10^{-6}-10^{-7}$ range.\\

\noindent {\bf REFERENCES}
\begin{enumerate}
\item A detailed review on the experimental and theoretical states of weak
radiative decays of hyperons is: J. Lach and P. Zenczykowski, Int.\ J.\ of
Mod.\ Phys.\ A (in press).
\vspace{-0.25cm}

\item P. Singer, in ``Puzzles on the Electroweak Scale'' (World Scientific,
Eds. Z. Ajduk, S. Pokorski and A.K. Wroblewski) p. 126 (1992).
\vspace{-0.25cm}

\item A.N. Kamal, Phys.\ Rev.\ {\bf D28}, 2176 (1983).
\vspace{-0.50cm}

\item T. Uppal and R.C. Verma, Phys.\ Rev.\ {\bf D47}, 2858 (1993).
\vspace{-0.25cm}

\item H-Y. Cheng et al., Phys.\ Rev.\ {\bf D51}, 1199 (1995).
\vspace{-0.25cm}

\item H.-Y. Cheng and B. Tseng, IP-ASTP-03-95 (1995).
\vspace{-0.25cm}

\item Particle Data Group, Phys.\ Rev.\ {\bf D50}, 1173 (1994).
\vspace{-0.25cm}

\item Ya.I. Kogan and M.A. Shifman, Sov.\ J.\ Nucl.
Phys.{\bf 38}, 628 (1983).
\vspace{-0.25cm}

\item P. Singer, Phys.\ Rev.\ {\bf D42}, 3255 (1990).
\vspace{-0.25cm}

\item M.A. Shifman, A.I. Vainshtein and V.I. Zakharov,
 Phys.\ Rev.\ {\bf D18}, 2853 (1978).
\vspace{-0.25cm}

\item N.G. Deshpande, P. Lo, J. Trampetic, G. Eilam
and P. Singer, Phys.\ Rev.\ Lett.\ {\bf 59}, 183 (1987);
S. Bertolini, F. Borzumati and A. Masiero, Phys.\ Rev.\ Lett.\
{\bf 59}, 180 (1987).
\vspace{-0.50cm}

\item A. Buras, M. Misiak, M. M\"{u}nz and S. Pokorski,
Nucl.\ Phys.\ B{\bf 424}, 374 (1994).
\vspace{-0.25cm}

\item F.M. Borzumati, Zeit.\ f.\ Physik C{\bf 63}, 291 (1994);
J. Hewett, SLAC-PUB-6521 (1994).
\vspace{-0.50cm}

\item F.J. Gilman and M.B. Wise, Phys.\ Rev.\ {\bf D19}, 976 (1979).
\vspace{-0.25cm}

\item A.N. Kamal and R.C. Verma, Phys.\ Rev.\ {\bf D26}, 190 (1982);
Lo Chong-Huah, Phys.\ Rev.\ {\bf D26}, 199 (1982).
\vspace{-0.25cm}

\item N.G. Deshpande, X.-G. Xe and J. Trampetic, Phys.\ Lett. (in press).
\vspace{-0.25cm}

\item G. Eilam, A. Ioannissian, R.R. Mendel and P. Singer, TECHNION-PH-95-18(199
\vspace{-0.25cm}

\item L. Bergstrom and P. Singer, Phys.\ Lett.\ {\bf 169B}, 297 (1986).
\vspace{-0.25cm}

\item R. Safadi and P. Singer, Phys.\ Rev.\ {\bf D37}, 697 (1988); {\bf D42}, 18
\vspace{-0.25cm}

\item T. Inami and C.S. Lim, Prog.\ Theor.\ Phys.\ {\bf 65}, 297 (1981).
\vspace{-0.25cm}

\item N.G. Deshpande and J. Trampetic,
Phys.\ Rev.\ Lett.\ {\bf 60}, 2583 (1988);
P.J. O'Donnell, M. Sutherland and H.K.K. Tung,
Phys.\ Rev.\ {\bf D46}, 4091 (1992).
\vspace{-0.25cm}

\item L. Bergstrom, R. Safadi and P. Singer, Zeit. f.\ Physik.\ C {\bf 37}, 281
\vspace{-0.25cm}

\item A.J. Buras, M. Jamin and M.E. Lautenbacher,
Nucl.\ Phys.\ B {\bf 408}, 209 (1993).
\vspace{-0.25cm}

\item G. Burdman, E. Golowich, J.L. Hewett and S. Pakvasa, SLAC-PUB-66921.
\vspace{-0.25cm}

\item S. Bertolini, M. Fabbrichesi and E. Gabrielli,
Phys.\ Lett.\ {\bf B327}, 136 (1994).
\vspace{-0.25cm}

\item I.F. Albuquerque et al.,  Phys.\ Rev.\ {\bf D50}, 18 (1994).
\vspace{-0.25cm}

\item
M. Ciuchini et al., Phys.\ Lett.\ B {\bf 316}, 127 (1993);
Nucl.\ Phys. B {\bf 415}, 403 (1994).
G. Cella et al., Nucl.\ Phys.\ B {\bf 431}, 417 (1994);
M. Misiak and M. M\"{u}nz, Phys.\ Lett.B {\bf 344}, 308 (1994).
\vspace{-0.25cm}

\item R. Ammar et al., Phys.\ Rev.\ Lett. {\bf 71}, 674 (1993).
\vspace{-0.25cm}

\item M.S. Alam et al., Phys.\ Rev.\ Lett. {\bf 74}, 2885 (1995).
\vspace{-0.25cm}

\item S. Timm et al., Phys.\ Rev.\ {\bf D51}, 4638 (1995).
\vspace{-0.25cm}

\item Y. Hara, Phys.\ Rev.\ Lett. {\bf 12}, 378 (1964); S.-Y. Lo,
Nuovo Cimento {\bf 37}, 753 (1965).
\vspace{-0.25cm}

\item N. Vasanti, Phys.\ Rev.\ {\bf D13}, 1889 (1976);
 L.-F. Li and Y. Liu, Phys.\ Lett.\ B {\bf 195}, 281 (1987);
M.K. Gaillard, Phys.\ Lett.\ B{\bf 211}, 189 (1988).
\vspace{-0.25cm}

\item A.N. Kamal and Riazuddin, Phys.\ Rev.\ {\bf D28}, 2317 (1983);
 P. Zenczykowski, Phys.\ Rev.\ {\bf D40}, 2290 (1989);
Int.\ J.\ Theor.\ Phys.\ {\bf 29}, 1327 (1990).
\vspace{-0.25cm}

\item F. Close and H.R. Rubinstein, Nucl.\ Phys.\ B{\bf 173}, 477 (1980);
K.G. Rauh, Z.\ Phys.\ C {\bf 10}, 81 (1981);
M.B. Gavela et al., Phys.\ Lett.\ {\bf 101B}, 417 (1981).
\vspace{-0.25cm}

\item W.F. Kao and H.J. Schnitzer,
Phys.\ Rev.\ {\bf D37}, 1912 (1988).
\vspace{-0.25cm}

\item P. Zenczykowski, Phys.\ Rev.\ {\bf D44}, 1485 (1981);
X.-L. Chen, C.-S. Gao and X.-Q. Li, Phys.\ Rev. {\bf D 51}, 2271 (1995).
\vspace{-0.25cm}

\item H. Neufeld, Nucl.\ Phys.\ {\bf B402}, 166 (1992);
E. Jenkins et al., Nucl.\ Phys. B{\bf 397}, 84 (1993).
\vspace{-0.25cm}

\item S.G. Kamath, Nucl.\ Phys. B{\bf 198}, 61 (1982); J.O. Eeg, Z.\ Phys.\ C
{\bf 21}, 253 (1984).
\vspace{-0.25cm}

\item J.J. Sakurai, ``Currents and Mesons'' (University of Chicago Press,
Chicago) 1969; N.M. Kroll, T.D. Lee and B. Zumino, Phys.\ Rev.\ {\bf 157},
1376 (1967).
\vspace{-0.25cm}

\item D. Palle, Phys.\ Rev.\ {\bf D36}, 2863 (1967).
\vspace{-0.25cm}

\item M. Gremm, F. Krueger and L.M. Sehgal, Phys.\ Lett.\ (in press).

\end{enumerate}

\end{document}